\documentclass[aps,prd,showpacs,preprint,amssymb,nofootinbib]{revtex4-1}

\usepackage{amssymb,amsmath}
\usepackage{amsfonts}
\usepackage{mathrsfs}
\usepackage{txfonts}
\usepackage{marvosym}
\usepackage{amssymb}

\begin{document}


\title[A uniqueness criterion for the Fock quantization of scalar fields]{A uniqueness criterion for the Fock quantization of scalar fields \\ with time dependent mass}

\author{Jer\'onimo Cortez}
\affiliation{Departamento de F\'\i sica,
Facultad de Ciencias, Universidad Nacional Aut\'onoma de
M\'exico, M\'exico D.F. 04510, Mexico.}
\email{jacq@ciencias.unam.mx}
\author{Guillermo A. Mena
Marug\'an${}^\star$, Javier Olmedo${}^\dagger$}
\affiliation{Instituto de Estructura de la Materia, IEM-CSIC,
Serrano 121, 28006 Madrid, Spain.}
\email{mena@iem.cfmac.csic.es${}^\star$, olmedo@iem.cfmac.csic.es${}^\dagger$}
\author{Jos\'e M. Velhinho}
\affiliation{Dept. de F\'{\i}sica, Universidade
da Beira Interior, R. Marqu\^es D'\'Avila e Bolama,
6201-001 Covilh\~a, Portugal.}
\email{jvelhi@ubi.pt}

\begin{abstract}
A major problem in the quantization of fields in curved spacetimes is the ambiguity in the choice of a Fock representation for the canonical commutation relations. There exists an infinite number of choices leading to different physical predictions. In stationary scenarios, a common strategy is to select a vacuum (or a family of unitarily equivalent vacua) by requiring invariance under the spacetime symmetries. When stationarity is lost, a natural generalization consists in replacing time invariance by unitarity in the evolution. We prove that, when the spatial sections are compact, the criterion of a unitary dynamics, together with the invariance under the spatial isometries, suffices to select a unique family of Fock quantizations for a scalar field with time dependent mass.
\end{abstract}


\maketitle

In contrast with the situation found in Quantum Mechanics (where one can appeal to the Stone-von Neumann uniqueness theorem \cite{simon}), linear canonical transformations (even if time independent) are not generally implemented as unitary transformations in Quantum Field Theory (QFT). This fact introduces an ambiguity in the choice of a Fock representation for the cannonical commutation relations (CCRs). Different choices of creation and annihilationlike variables are related by linear canonical transformations which cannot be all unitary under quantization. As a consequence, descriptions which are classically equivalent become inequivalent in the quantum realm. The number of linear canonical transformations that cannot be implemented unitarily is infinite, actually. Therefore, there exist infinitely many (intrinsically) distinct possibilities for the choice of creation and annihilation operators, and hence of inequivalent vacua \cite{wald}. This is a severe problem, since each of these Fock representations provides a QFT with different physical predictions.

In order to remove this ambiguity, inherent to QFT in curved spacetimes, the usual strategy is to employ the spacetime symmetries of the field theory. Namely, these classical symmetries must be inherited in the quantum theory, where they must have a natural unitary implementation. This is more stringent than just unitarity: one demands that the structures used in the Fock quantization, and hence the vacuum, are invariant under these symmetries. This immediately ensures unitarity, while the opposite is not generally true (unitarity does not require vacuum invariance).

The relevant information on the choice of creation and annihilationlike variables (up to redundancies like the mixing of only creation variables) is encoded in a basic structure called the {\it complex structure}. Let us consider the symplectic form of our field system, which encapsules all the information about the Poisson brackets. Then, a complex structure is a linear map in phase space whose square is minus the identity, leaves invariant the symplectic form and, combined with it, provides an inner product in that phase space. Given a complex structure, there is a standard procedure (see \cite{wald}) to obtain a Fock representation of the CCR's (or, more precisely, of the corresponding Weyl relations). Now, we say that a certain group of symmetries admits a natural unitary implementation in a Fock quantization if and only if the chosen complex structure is invariant under that group.

The best known example where a unique vacuum is selected by imposing symmetries is QFT in Minkowski spacetime: uniqueness follows from the invariance under the Poincar\'e group \cite{wald}. But in general cases, and in particular in nonstationary settings, there is simply not sufficient symmetry to pick up a unique Fock representation.

In the absence of stationarity, it seems quite natural to minimally relax the requirement of
invariance under time evolution, replacing it with the condition that the dynamics be unitary. This is precisely the criterion, additional to symmetry invariance, that we will put forward in this letter to reach uniqueness in the Fock quantization. In fact, this criterion has been successfully tested in the context of inhomogeneous cosmologies \cite{unique-gowdy-1,BVV2} as well as for scalar fields with generic time dependent mass on $d$-spheres, with $d=1,2,3$ \cite{CMV8} (see also \cite{jackiw} for different criteria concerning free fields in 1+1 dimensional de Sitter space).

Specifically, we are going to consider the Fock quantization of a scalar field with generic time varying mass (i.e., subject to a time dependent quadratic potential) on a general Riemannian compact space in three or less (spatial) dimensions. We will prove that a unique, preferred Fock representation is selected by imposing the criterion of unitary dynamics and a natural unitary implementation of the spatial symmetries.

It is worth commenting that scalar fields propagating in certain nonstationary spacetimes can be reformulated, via a time dependent canonical transformation in which the field is scaled by a time function, as scalar fields with a time varying mass but in a static background. For instance, this occurs in Friedmann-Robertson-Walker spacetimes. Hence, our analysis has immediate applications in cosmology, e.g. in the study of cosmological perturbations in models with compact spatial topology. This includes the physically important case of flat models with compact sections of  three-torus topology.

The description of the scalar field as a Klein-Gordon field with time dependent mass is known to have particularly nice properties\footnote{
Unitary dynamics has in fact been achieved in some particular cases, though the uniqueness of the representation respecting this unitarity has not been proved before to the best of our knowledge. A detailed discussion of the relation between the approach presented here
and earlier work on QFT in Friedmann-Robertson-Walker spacetimes will be the subject of a future publication.}
\cite{weiss}, and recent results indicate that it can have a privileged behavior in terms of its quantum dynamics \cite{zejaguije}. Moreover, it is precisely the field description adopted in Mukhanov's formalism for cosmological perturbations \cite{mukhanov}. In this context, what we are going to show is that, rather than renouncing to a unitary dynamics in cosmology, one can select a unique Fock representation precisely by invoking
it, together with the spatial symmetries.

In more detail, we will analyze a real scalar field $\varphi$ propagating in a globally hyperbolic backgroud $\mathbb{I} \times \Sigma$, where $\mathbb{I}$ is a time interval and $\Sigma$ is a Riemannian compact manifold with metric $h_{ab}$. We will study only the case of orthogonal foliations. To pass to the canonical formulation, we choose an (arbitrarily) fixed time $t_0$, and consider the pairs of data
$(\varphi,P)=(\varphi_{|t_0},\sqrt{h}{\dot\varphi}_{|t_0})$, where $\dot \varphi$ denotes the time derivative and
$h$ is the determinant
of $h_{ab}$. The phase space is thus the set of pairs $\{(\varphi,P)\}$, equipped with the
symplectic form $\Omega$ determined by the Poisson brackets $\{\varphi(t_0,x),P(t_0,y)\}=\delta(x-y)$
(independent of the choice of $t_0$).
Note that the configuration variable $\varphi$ is a scalar and the momentum $P$ is a scalar density.

We introduce now the Laplace-Beltrami (LB) operator $\Delta$ associated with the metric $h$, as well as
the complex structure:
\begin{eqnarray}
\label{cano-cs} J_0\left( \begin{array}{c} \varphi \\
P\end{array}\right) = \left( \begin{array}{cc} 0 &
-(-h \Delta)^{-1/2}
\\(-h \Delta)^{1/2} &  0\end{array}\right) \left(
\begin{array}{c} \varphi \\ P\end{array}\right).
\end{eqnarray}
Clearly, the Fock representation defined by $J_0$ is the analogue of the free massless field representation for
Minkowski spacetime. In that case, one can see that $J_0$ is invariant under the evolution, and therefore a
natural unitary implementation of the dynamics is achieved then.

Let us consider the general case of a field with time dependent mass,
whose canonical equations of motion are
\begin{equation}
\label{fieldequations}
 \dot\varphi:=\frac{1}{\sqrt{h}}P, \quad
\dot P=\sqrt{h}[\Delta\varphi -f(t)\varphi].
\end{equation}
Here $f(t)$ is a rather arbitrary
function (apart from some mild
conditions specified in \cite{CMV8}). This set of equations is equivalent to
a Klein-Gordon field equation with square mass equal to $f(t)$.

We next discuss the necessary and sufficient conditions for a unitary implementation of the dynamics dictated by \eqref{fieldequations}, with respect to the Fock representation determined by the complex structure $J_0$.
Crucial in this analysis are the properties of the LB operator
in compact spaces \cite{compact}, which allow us to perform a mode decomposition of the field,
just like with the standard Fourier series. In this general setting, the inner product
on the space of functions on $\Sigma$ is determined by the integration with respect to the
metric volume element.

Let then $\{\Psi_{nl_n}\}$ denote a complete set of real orthonormal eigenfunctions of the
operator $\Delta$, corresponding to the discrete set of eigenvalues $\{-\omega^2_n\}$,
which satisfy $\omega^2_n\to\infty$ as $n\to\infty$.
In general, there may be some degeneracy, i.e., the eigenspace
with eigenvalue (minus) $\omega^2_n$ may have dimension $g_n$
greater than the unity (although necessarily finite).
This degeneracy is accounted for by the label $l_n=1,...,g_n$.
In what follows, every sum is performed over the whole spectrum of eigenvalues,
including degeneracy.

We then decompose the field $\varphi$ in terms of eigenmodes:
\begin{equation}
\varphi=\sum_nq_{nl_n}\Psi_{nl_n}.
\end{equation}
The
degrees of freedom of the system are thus encoded in a discrete set of
real modes $q_{nl_n}$, which obey decoupled equations of
motion
\begin{equation} \label{q-eq} \ddot
q_{nl_n}+[\omega_n^2+f(t)]q_{nl_n}=0.
\end{equation}
Notice that all the modes with
the same value of $n$ have the same dynamics,
regardless of the degeneracy label $l_n$.

The unitary implementation of the dynamics depends on the behavior
of the canonical evolution matrices when $n\to\infty$.
It is convenient to introduce the
annihilationlike variables
\begin{equation}
\label{basic-var} a_{nl_n}={\frac{1}{\sqrt{2\omega_n}}} \left(\omega_nq_{nl_n}
+i p_{nl_n}\right),
\end{equation}
together with their complex conjugates: the
creationlike variables $a_n^*$ (clearly, this definition is meaningless for $\omega_n=0$;
however, our analysis is not affected by any finite number of modes, so we will
consider exclusively nonzero modes from now on).
Here, $p_{nl_n}={\dot q}_{nl_n}$ is the canonical momentum conjugate
to $q_{nl_n}$.   In these variables, the complex structure
$J_0$, which has already been diagonalized in $2\times 2$ blocks by means of the mode
decomposition, is further diagonalized within each block, taking the standard
form $J_0(a_{nl_n})=i a_{nl_n}$ and $J_0(a_{nl_n}^{\ast})=-i a_{nl_n}^{\ast}$ (this means that $a_{nl_n}$ and
$a_{nl_n}^{\ast}$ are precisely the variables quantized as the annihilation and creation
operators in the corresponding Fock representation).

The time evolution of the canonical variables $(a_{nl_n},a_{nl_n}^*)$, from the fixed reference time $t_0$
to an arbitrary time $t$, is a linear transformation which is block diagonal, owing to the decoupling of modes,
and which has the form
\begin{equation}
\label{bogo-transf} a_{nl_n}(t)=\alpha_n(t,t_0)a_{nl_n}(t_0)+ \beta_n(t,t_0)a_{nl_n}^*(t_0),
\end{equation}
given its independence on the degeneracy label.
For each label $n$, the functions $\alpha_n(t,t_0)$ and $\beta_n(t,t_0)$,
which completely characterize the classical evolution, satisfy
\begin{equation}
\label{symp}
|\alpha_n(t,t_0)|^2 -|\beta_n(t,t_0)|^2=1,\qquad  \forall t,t_0.
\end{equation}

As explained in \cite{unit-gt3}, the canonical transformation \eqref{bogo-transf} is unitarily
implementable in the Fock representation defined by the complex structure $J_0$ if and only if
\begin{equation}
\label{sqsb}
\sum_n g_n |\beta_n(t,t_0)|^2<\infty.
\end{equation}
Often, this condition is presented
as the requirement that the evolved vacuum contain only a finite number of particles.

We can now make use of the asymptotic analysis performed in \cite{CMV8} (see Sec III.B),
for the harmonic oscillator equation of type (\ref{q-eq}), in the limit of large $\omega_n^2$
(i.e., $n\to\infty$).
It was shown that, for any function $f(t)$, the leading term in
$\beta_n(t,t_0)$ is proportional to $1/\omega_n^2$.
Therefore, condition \eqref{sqsb} is equivalent to
\begin{equation}
\label{sqso}
\sum_n\frac{g_n}{\omega_n^4}<\infty.
\end{equation}
Actually, this last condition is satisfied for all Riemannian compact manifolds in three or less dimensions. One can see this
from well known asymptotic properties of the spectrum of the LB operator, which ensure that the number of eigenstates with eigenvalue
equal or smaller in norm than $\omega^2$ - known as the {\it counting function} - does not grow
faster than $\omega^{d}$ \cite{compact}.

Proceeding with our analysis, suppose that the manifold $(\Sigma,h_{ab})$ possesses an isometry group $G$. Then the LB operator
is invariant under $G$, and it follows immediately that this group of transformations translates into symmetries
of the field equations (\ref{fieldequations}), or into canonical transformations which
commute with the dynamics in the canonical formulation. Consequently,
we require those symmetries to have a natural unitary implementation in the quantum theory. This is
automatically granted in the Fock representation defined by the complex structure $J_0$, since it depends only on the metric,
and is therefore invariant under the group $G$.

Hence, we have a complex structure - $J_0$ - which is invariant under any existing symmetry group and provides a representation
in which the dynamics dictated by (\ref{q-eq}) is unitary.
We are now going to show that there is no other inequivalent
Fock representation with the same properties. Namely, we will see that
any other complex structure which is invariant under a symmetry group $G$ and allows
for a unitary implementation of the dynamical evolution, defines a quantum representation
which is unitarily equivalent to the one fixed by $J_0$.

Let us start with the description of $G$-invariant complex structures, based on a simple
application of Schur's lemma \cite{unique-gowdy-1,BVV2,CMV8}.
First, we consider the action of the group $G$ on the set of fields $\varphi$,
i.e. on the configuration space, hereafter called $\cal Q$. Since $G$ preserves the metric,
this action is naturally unitary (with respect to the inner product on the space of
functions $\cal Q$). Each eigenspace of $\Delta$
(corresponding to a given eigenvalue) is itself a representation of $G$,
because  $G$ commutes with $\Delta$.
Since the action is unitary, a given eigenspace is either
irreducible under $G$ or can be decomposed into a set of mutually orthogonal irreducible
subspaces. Then, in the natural decomposition of $\cal Q$ as a direct sum
of (finite dimensional) eigenspaces ${\cal Q}^{\omega_n}$ of $\Delta$, each eigenspace
can be decomposed in turn as a direct sum of irreducible representations ${\cal Q}^{m_n}$ of $G$.
Obviously, the sum of the dimensions $g_{m_n}$ of all such representations ${\cal Q}^{m_n}$
must equal the degeneracy $g_n$,
and since the representations are at least one-dimensional, we have $1 \leq g_{m_n} \leq g_n$.

We next consider the analogous decompositions of the space $\cal P$ of momentum
fields $P$ (here, the integration is performed with the inverse volume element).
Putting then configuration and momentum fields together, we obtain the corresponding
decompositions of the phase space $\Gamma$:
\begin{equation}
\label{irre0}
{\Gamma}=\bigoplus_n {\Gamma}^{\omega_n}= \bigoplus_{n,m_n} {\Gamma}^{m_n},
\qquad {\Gamma}^{m_n}=  {\cal Q}^{m_n}\oplus {\cal P}^{m_n}.
\end{equation}
Taking into account that the group $G$ acts in the same way on fields $\varphi$ and on momentum
fields $P$, we easily realize that its action coincides on the irreducible spaces
${\cal Q}^{m_n}$ and ${\cal P}^{m_n}$.

As a first application of Schur's lemma \cite{schur}, we conclude that every
$G$-invariant complex structure is block diagonal with respect to the decomposition
(\ref{irre0}), since invariant transformations do not mix different irreducible representations.
Thus, an invariant complex structure $J$ must have the form
\begin{equation}
\label{j}
{J}=\bigoplus {J}_{m_n},
\end{equation}
where each $J_{m_n}$ is an invariant complex structure on the corresponding space $\Gamma^{m_n}$.
For each of these spaces, it is always possible to find a basis formed by configuration
variables $q_{nl_n}$ and corresponding momenta $p_{nl_n}$, associated with a subset of orthonormal eigenmodes
as those introduced above (recall that the different spaces ${\cal Q}^{m_n}$ are orthogonal to each other).
For a given $n$, the complete set $\{q_{nl_n}, p_{nl_n}\}$ is formed by the union
of all such subsets, when all the subspaces $\Gamma^{m_n}$ of $\Gamma^{\omega_n}$ are taken into account.
To each $J_{m_n}$, then, there corresponds a matrix characterized by four square
blocks: $J^{qq}_{m_n}$, $J^{qp}_{m_n}$, $J^{pq}_{m_n}$, and
$J^{pp}_{m_n}$.
The invariance conditions on those blocks become invariance
conditions for matrices in the given irreducible representation of $G$. Hence, again by Schur's
lemma, each of the above matrices must be proportional to the identity ${\bf I}$, i.e.,
\begin{equation}  \hspace*{-2pc}
J^{qq}_{m_n}=a_{m_n} {\bf I},
\quad J^{pq}_{m_n}=b_{m_n} {\bf I},  \quad J^{qp}_{m_n}=c_{m_n} {\bf I}\quad {\rm and }\quad J^{pp}_{m_n}=d_{m_n} {\bf I},
\end{equation}
where $a_{m_n}$, $b_{m_n}$, $c_{m_n}$,
and $d_{m_n}$ are real numbers.
It follows that  invariant complex structures further decompose within each sector
$\Gamma^{m_n}$ into a block-diagonal form such that each block is given by the same matrix,
namely, a 2-dimensional complex structure which only mixes $q_{nl_n}$ with $p_{nl_n}$ for each value of $l_n$.

To proceed further, it is convenient to switch again from the real basis $\{q_{nl_n}, p_{nl_n}\}$ to the complex variables $a_{nl_n}$ and $a_{nl_n}^{\ast}$, and relate a general invariant complex structure with $J_0$.
One can easily show that this relation is always of the form $J=K J_0 K^{-1}$, where $K$ is a symplectic transformation.
Given our above description of invariant complex
structures, the symplectic transformation $K$ can also be decomposed in $2\times 2$ blocks that are all identical
in each space $\Gamma^{m_n}$. So, the
information necessary to determine a given invariant complex
structure is encoded in a discrete set of $2\times 2$ matrices
${\cal K}_{m_n}$, one per each irreducible representation of $G$.
They are then totally determined by a set $\{\kappa_{m_n},\lambda_{m_n}\}$ of complex numbers
that satisfy the condition $|\kappa_{m_n}|^2-|\lambda_{m_n}|^2=1$, $\forall m_n$ (see e.g. \cite{CMV8}).

On the other hand, it is known that any symplectic transformation $R$ admits a
unitary implementation with respect to the complex
structure $J=K J_0 K^{-1}$ if and only if the
transformation $K^{-1}RK$ is unitarily implementable
with respect to $J_0$ \cite{unique-gowdy-1}. Therefore, the condition for unitary implementation
of the dynamics with respect to a complex structure $J=K J_0 K^{-1}$ can be replaced by
a condition with respect to $J_0$.
The effect of the transformation $K$ is to replace the functions $\alpha_n$ and $\beta_n$ by new ones, and to possibly lift part of the degeneracy because, in addition to the value of $n$, these new functions can depend
on the irreducible representation of $G$ as well.
In particular, one can compute the relation between the new beta coefficients, hereafter denoted by
$\beta^J_{m_n}$, and the functions $\alpha_n$ and $\beta_n$:
\begin{equation}  \hspace*{-2pc}
\beta^J_{m_n}(t,t_0)=(\kappa_{m_n}^*)^2\beta_n
(t,t_0)-\lambda_{m_n}^2\beta^*_n(t,t_0)+2 i
\kappa_{m_n}^*\lambda_{m_n} {\Im}[\alpha_n(t,t_0)], 
\end{equation}
where the symbol $\Im$ denotes the imaginary part.

Let us then suppose that, besides $J_0$, there exists another invariant complex structure $J$ which also
allows for a unitary implementation of the dynamics. This unitarity implies that
the sequences given by $\sqrt{g_{m_n}}\beta^J_{m_n}(t,t_0)$ (where we have taken into account the
remaining degeneracy)
are square summable in $m_n$ at all possible values of $t$. We are then in a situation which is completely similar to that
studied in \cite{CMV8}, and our result is readily obtained by repeating
the arguments presented in that work. First,
using the fact that condition
(\ref{sqsb}) is satisfied, it follows from the square summability of $\sqrt{g_{m_n}}\beta^J_{m_n}(t,t_0)$ that
the sequences with elements
$\sqrt{g_{m_n}}\,{\Im}[\alpha_n(t,t_0)]\, \lambda_{m_n}/\kappa_{m_n}^*$
are also square summable, $\forall t\in \mathbb{I}$.
Moreover, the asymptotic behavior of the sequences $\left\{{\Im}[\alpha_n(t,t_0)]\right\}$ was analyzed in \cite{CMV8}
(see Secs. IIIB and IVC), and it was shown
there that square summability follows as well for the sequence
\begin{equation}
\label{43} \left\{
\,\sqrt{g_{m_n}}\frac{\lambda_{m_n}}{\kappa_{m_n}^*}\,\sin{\left[
\omega_n(t-t_0)+\int_{t_0}^t
d\bar{t}\frac{f(\bar{t})}{2\omega_n}\right]}\right\}.
\end{equation}

Finally, we can apply the line of reasoning used in Sec. IVC of \cite{CMV8}
--employing Luzin's theorem and integrating over a suitable measurable
set in the time interval $\mathbb{I}$-- to
show that the square summability of (\ref{43}) implies that $\sum_{n,m_n} g_{m_n}|\lambda_{m_n}|^2<\infty$.
But it can be easily seen \cite{unique-gowdy-1,CMV8} that this is precisely the necessary and sufficient condition for the unitary equivalence between the Fock representation defined by $J$ and the corresponding one defined by $J_0$. This concludes our proof.

Summarizing, we have shown that the Fock representation determined by $J_0$, analog of the free massless field representation for Minkowski spacetime, respects the invariance under the spatial isometries and provides a unitary dynamics for any compact spatial manifold in three or less dimensions. Furthermore, among the class of representations defined by invariant complex structures, this choice is indeed the unique one (up to unitary equivalence) with a unitary evolution.

This uniqueness result finds applications in a wide class of systems, including test scalar fields with time dependent potentials or, in much more realistic scenarios, (conveniently scaled) cosmological perturbations evolving in a nonstationary, homogeneous spacetime, e.g. in Friedmann-Robertson-Walker cosmologies. Our results are valid for perfect fluids with isotropic perturbations of the energy-momentum tensor in this class of universes. Furthermore, they can be straightforwardly extended to tensor perturbations. Besides, preliminary calculations indicate that the proposed uniqueness criterion is valid even for spinorial fields, which can be treated similarly without introducing radically new conceptual difficulties.

It is worth noting that, in intricate situations where no clear symmetries can be identified
in the spatial manifold $\Sigma$, one can still resort to unitary groups in the space of
square integrable functions on $\Sigma$, constructed from operators that leave
invariant the eigenspaces of the LB operator. This more general notion of symmetry, however, may
force the unitary implementation of transformations that arise just from an accidental degeneracy
in the spectrum of the LB operator.

We have addressed the uniqueness problem taking for granted a certain choice of field description. A complete analysis of the freedom permitted by linear canonical transformations should also account for changes of field description obtained by means of time dependent scalings, which seem natural to contemplate for fields propagating in nonstationary backgrounds or in cases where no fundamental field is fixed from scratch.  Concerning this freedom, fortunately, the results of \cite{zejaguije} strongly support that this ambiguity is also removed by the criterion of symmetry invariance and unitary dynamics.

The proof that the Fock representation determined by $J_0$ implements the field dynamics in a unitary way depends critically on the dimension of the Riemannian manifold. In general, condition (\ref{sqso}) is not fulfilled in four or more spatial dimensions (the inverse of the LB operator in the orthogonal complement of the zero modes is not Hilbert-Schmidt). In such cases, an open issue is whether one can still find a Fock representation which leads to a unitary evolution and analyze whether its equivalence class is picked up uniquely by our criterion. Finally, the assumption of spatial compactness is essential to avoid infrared problems \cite{wald}. However, in cosmology, where scales beyond those provided by the Hubble radius are expected to play no significant physical role, the infrared behavior can be ignored, and there is no fundamental difference between the treatment of compact and noncompact scenarios. In this context, therefore, our uniqueness result has complete generality in 3+1 dimensions.

\section*{Acknowledgements}
This work was supported by the research grants MICINN FIS2008-06078-C03-03 and
CPAN CSD2007-00042 from Spain, DGAPA-UNAM
IN108309-3 from Mexico, CERN/FP/116373/2010 and PTDC/CTE-ATM/73607/2006 from Portugal. 
J.O. acknowledges CSIC by financial support under the grant JAE-Pre\_08\_00791.

\bibliographystyle{plain}

\end{document}